\documentclass[12pt]{article} 
\usepackage{epsfig}
\usepackage{a4}
\usepackage{latexsym}
\usepackage{cite}

\usepackage{pslatex}
\usepackage[latin1]{inputenc}
\usepackage[T1]{fontenc}

\newcommand{\hspn}{{\hspace{-4mm}}}

\newcommand{\beq}{\begin{equation}}
\newcommand{\eeq}{\end{equation}}
\newcommand{\bea}{\begin{eqnarray}}
\newcommand{\eea}{\end{eqnarray}}
\newcommand{\nn}{\nonumber}
\newcommand{\MSb}{$\overline{\mbox{MS}}$}
\newcommand{\as}{\alpha_{\rm s}}
\newcommand{\ar}{a_{\rm s}}
\newcommand{\ra}{\rightarrow}
\newcommand{\ep}{\epsilon}

\newcommand{\ec}{\gamma_{\rm e}}
\newcommand{\ecs}{\gamma_{\rm e}^{\:2}}
\newcommand{\ect}{\gamma_{\rm e}^{\:3}}
\newcommand{\ecf}{\gamma_{\rm e}^{\:4}}

\begin{document}
\setlength{\parskip}{0.15cm}
\setlength{\baselineskip}{0.53cm}

\def\Fh{{F^{\:\! h}}}
\def\Qs{{Q^{\, 2}}}
\def\qs{{q^{\:\! 2}}}
\def\Fone{{F_{\:\! 1}}}
\def\Ftwo{{F_{\:\! 2}}}
\def\FL{{F_{\:\! L}}}
\def\F3{{F_{\:\! 3}}}
\def\A4{{A_4}}
\def\DDi{{{\cal D}_{i}}}
\def\DDk{{{\cal D}_{k}}}
\def\DD#1{{{\cal D}_{#1}^{}}}
\def\x1{{(1 \! - \! x)}}
\def\z#1{{\zeta_{\:\! #1}}}
\def\zss{{\zeta_{2}^{\,2}}}
\def\zst{{\zeta_{3}^{\,2}}}
\def\zts{{\zeta_{2}^{\,3}}}
\def\ca{{C^{}_A}}
\def\cas{{C^{\: 2}_A}}
\def\cat{{C^{\: 3}_A}}
\def\cf{{C^{}_F}}
\def\cfs{{C^{\: 2}_F}}
\def\cft{{C^{\: 3}_F}}
\def\cff{{C^{\: 4}_F}}
\def\nf{{n^{}_{\! f}}}
\def\nfs{{n^{\,2}_{\! f}}}
\def\nft{{n^{\,3}_{\! f}}}
\def\dabc2{{d^{\:\!abc}d_{abc}}}
\def\dabcnc{{{d^{abc}d_{abc}}\over{n_c}}}
\def\fl11{fl_{11}}
\def\b#1{{{\beta}_{#1}}}
\def\bb#1#2{{{\beta}_{#1}^{\,#2}}}

\begin{titlepage}
\noindent
DESY 09-129 \hfill {\tt arXiv:0908.2746 [hep-ph]}\\
SFB/CPP-09-74 \\
LTH 839 \\[1mm]
August 2009 \\
\vspace{1.5cm}
\begin{center}
\LARGE
{\bf Higher-order threshold resummation\\[1mm]
     for semi-inclusive $e^+e^-$ annihilation}

\vspace{2.5cm}
\large
S. Moch$^{\, a}$ and A. Vogt$^{\, b}$\\
\vspace{1.5cm}
\normalsize
{\it $^a$Deutsches Elektronensynchrotron DESY \\
\vspace{0.1cm}
Platanenallee 6, D--15738 Zeuthen, Germany}\\
\vspace{0.5cm}
{\it $^b$Department of Mathematical Sciences, University of Liverpool \\
\vspace{0.1cm}
Liverpool L69 3BX, United Kingdom}\\[1.2cm]
\vfill
\large
{\bf Abstract}
\vspace{-1mm}
\end{center}
The complete soft-enhanced and virtual-gluon contributions are derived for the
quark coefficient functions in semi-inclusive $e^+e^-$ annihilation to the
third order in massless perturbative QCD. These terms enable us to extend the 
soft-gluon resummation for the fragmentation functions by two orders to the 
next-to-next-to-next-to-leading logarithmic (N$^{\,3}$LL) accuracy. The 
resummation exponent is found to be the same as for the structure functions in 
inclusive deep-inelastic scattering. This finding, together with known results 
on the higher-order quark form factor, facilitates the determination of all 
soft and virtual contributions of the fourth-order difference of the 
coefficient functions for these two processes. Unlike the previous 
(N$^{\,2}$LL) order in the exponentiation, the numerical effect of the 
N$^{\,3}$LL contributions turns out to be negligible at LEP energies.

\vfill
\end{titlepage}

% ----------------------------------------------------------------------------

Semi-inclusive $e^+e^-$ annihilation (SIA) via a virtual photon or $Z$-boson, 
$e^+e^- \to\, \gamma/Z \to\, h +\! X$, is a classic process probing Quantum 
Chromodynamics (QCD), the theory of the strong interaction. A wealth of precise
measurements have been performed, at various center-of-mass (CM) energies
$\sqrt{s}$, of the total fragmentation function
\beq
\label{sigmaFtot} \qquad
  \frac{1}{\sigma_{\rm tot}} \: \frac{d \sigma^{\,h}}{dx}
   \;\: = \;\: \Fh(x,\Qs) \:\: ,
\eeq
where $h$ stands for a specific hadron species or the sum over all (charged) 
light hadrons, see Ref.~\cite{PDG08} for a general overview. In the CM frame
the scaling variable $x$ is the fraction of the beam energy carried by the 
hadron $h$, and $\Qs = s$ is the square of the four-momentum $q$ of the 
intermediate gauge boson. In perturbative QCD, the total (angle-integrated) 
fragmentation function $F_I^h \,\equiv\, \Fh\!$, as well as the transverse 
($F_T$), longitudinal ($F_L$) and asymmetric ($F_A$) fragmentation functions 
for the double-differential cross section 
$\,d\sigma^{\,h}/ dx\, d \cos\theta_h$ \cite{NasonW94}, are given by
\beq
\label{FapQCD}
  F_a^{\:\! h}(x,\Qs) \;\: = \; \sum_{\rm f = q,\,\bar{q},\,g} \;
  \int_x^1 {dz \over z} \; C^{}_{a,{\rm f}} \left( z,\as (\Qs) \right)
  \,  D_{\rm f}^{\,h} \bigg( \,{x \over z},\Qs\bigg)
  \; + \; {\cal O} \!\left( \frac{1}{Q} \right) 
  \:\: .
\eeq
Here $D_{\rm f}^{\,h}$ are the parton fragmentation functions, the final-state
(timelike, $\Qs = \qs$) analogue of the initial-state (spacelike, $\Qs = -\qs$)
parton distribution functions in deep-inelastic scattering (DIS).
Without loss of information in the present context, the renormalization scale
of $\as$ and the factorization scale of $D_{\rm f}^{\,h}$ have been identified
with the physical hard scale $\Qs$ in Eq.~(\ref{FapQCD}).
The coefficient functions $C^{}_{a,\rm f\,}$ are defined via expansions in the
strong coupling $\ar \,\equiv\, \as/(4\pi)$. 

Here we are interested in the dominant (anti-)$\,$quark contributions to 
$F_I^h$, $F_T^{\:\!h}$ and $F_A^{\:\!h}$,
\beq
\label{Cqexp}
  C^{}_{a,q} ( x,\as ) \;\: = \;\: \sigma_{\rm ew} \big(  
  \delta \x1 \:+\: 
  \ar\, c_{a,q}^{(1)}(x) \:+\: 
  \ar^{\,2} \, c_{a,q}^{(2)}(x) \:+\:
  \ar^{\,3} \, c_{a,q}^{(3)}(x) \:+\:
  \ldots \big) \:\: .
\eeq
The electroweak prefactors $\sigma_{\rm ew}$ can be found in Ref.~\cite
{NasonW94}. The first- and second-order coefficient functions have been 
calculated long ago in Refs.~\cite{NLOcoeff} and \cite{RvN2loop}, respectively.
More recently the latter results have been confirmed (and some typos corrected)
in two independent ways in Refs.\cite{MMV1,MMoch06}. The three-loop corrections
$c_{a}^{(3)}(x)$ have not been derived so far.

The coefficient functions in Eq.~(\ref{Cqexp}) include large-$x$ (threshold)
double-logarithmic enhancements of the form $\ar^{\,n}\: \x1^{-1} \ln^{\, k} 
\! \x1$ with $\,k = 0,\, \ldots, 2n-1$. Such contributions, which spoil the
convergence of the perturbation series at sufficiently large values of $x$,
can be resummed by the soft-gluon exponentiation \cite{SoftGlue1,SoftGlue2}.
For the process at hand this resummation has been worked out to next-to-leading 
logarithmic (NLL) accuracy in Ref.~\cite{CacCat01}. The inclusion of this
resummation has led to improvements in a recent global fit of fragmentation 
functions \cite{AKK08}. Hence an extension of the soft-gluon exponentiation 
for $e^+e^- \to\, \gamma/Z \to\, h +\! X$ to a higher accuracy is not only of
theoretical but also of phenomenological interest.

In this letter we employ the analytic continuation approach of Ref.~\cite{MMV1}
to derive the soft and virtual contributions to the third-order coefficient
functions in Eq.~(\ref{Cqexp}). 
These results are then used to extend the results of Ref.~\cite{CacCat01} 
to the next-to-next-to-next-to-leading logarithmic (N$^3$LL) accuracy reached 
before for inclusive deep-inelastic scattering \cite{MVV7} and the total cross 
sections for lepton-pair and Higgs-boson production in proton--(anti-)proton
collisions \cite{MV1,NNNLL05}. A substantial intermediate step towards
the present extension has been taken before in Ref.~\cite{Blumlein:2006pj}.
\pagebreak

% ----------------------------------------------------------------------------

Up to small contributions from higher-order group invariants entering at the 
third and higher orders,
the soft plus virtual contributions are the same for the DIS quark coefficient 
functions for $\Fone$, $\Ftwo$ and $\F3$ \cite{MVV6,MVV10}. The same holds for 
the corresponding (in this order) SIA coefficient functions for 
$F_{\:\!T}$, $F_{\:\!I}$ and $F_A$. Hence we will drop the index $a$ from now
on, and refer to the former coefficient functions collectively as 
$c_{\rm S}^{(l)}(x)$, and the latter as $c_{\rm T}^{(l)}(x)$.

In this limit the bare (unrenormalized and unfactorized) partonic DIS 
(spacelike) structure function $F_S^b$ is given by 
\cite{FF2vN,MVV8}
\beq
\label{eq:F1b}
  F_{\rm S}^{\,\rm b} (\as^{\,\rm b}, \Qs) \;\: = \;\:
  \delta(1-x) \:  + \: \sum_{l=1} ( \as^{\,\rm b\:\!} )^l
  \left( \Qs \over \mu^2 \right)^{\! -l\ep} F_{{\rm S},l}^{\,\rm b} 
\eeq
with
\bea
  & & \nn \\[-9mm]
\label{FSdec}
F_{\rm S,1}^{\,\rm b}
     &\!=\!& 2 {\cal F}_1\,\delta(1-x) + {\cal S}_{\,1} \nn \\[0.5mm]
F_{\rm S,2}^{\,\rm b}
     &\!=\!& ( 2 {\cal F}_2
           + \left({\cal F}_1\right)^2 ) \, \delta(1-x)
           + 2 {\cal F}_1 {\cal S}_{\,1} + {\cal S}_{\,2} \nn \\[0.5mm]
F_{\rm S,3}^{\,\rm b}
     &\!=\!& ( 2 {\cal F}_3
           + 2 {\cal F}_1 {\cal F}_2 ) \, \delta(1-x)
           + (\, 2 {\cal F}_2 + \left({\cal F}_1\right)^2 \,)\, {\cal S}_{\,1}
           + 2 {\cal F}_1 {\cal S}_{\,2} + {\cal S}_{\,3}
\:\: .
\eea
Here $\mu$ is the scale of dimensional regularization with $D = 4 - 2\ep$,
and $\ar^{\,\rm b}$ the bare strong coupling. ${\cal F}_l$ represents the
$l$-loop quark form factor 
\cite{FF2vN,MVV8,KL87,Gehrmann:2005pd,MVV9,Baikov:2009bg}.
The $x$-dependence of the real-emission functions ${\cal S}_k$ is given by the 
$D$-dimensional +-distributions 
\beq
\label{Dplus}
 f_{k\epsilon}(x) \; = \; [\,(1-x)^{-1-k\epsilon}\,]_+
                  \; = \; - {1 \over k\epsilon}\, \delta(1-x) + \sum_{i=0}\,
                        {(-k\epsilon)^i  \over i\, !}\, \DDi
\eeq
where we have introduced the abbreviation
$
  \DDk \: = \: [ \x1^{-1} \ln^{\,k}\! \x1 ]_+ 
$.

The transition to the bare SIA (timelike) fragmentation functions 
$F_{\rm S}^{\,\rm b}$ is performed as follows: 
In Eq.~(\ref{FSdec}) the factors $\,2 {\cal F}_l$ are replaced everywhere by $2\, {\rm Re} 
{\cal F}^{\!T}_l$ and all products ${\cal F}_k {\cal F}_l$ by 
$| {\cal F}^{\!T}_k {\cal F}^{\!T}_l |$, where ${\cal F}^{\!T}_l$ is the 
complex $l$-loop timelike form factor which can be obtained from the
spacelike ${\cal F}_l$ by Eq.~(3.3) of Ref.~\cite{FF2vN}.
The analytic continuation of the real-emission terms ${\cal S}_{\:\!k}$ is 
carried out as discussed in Ref.~\cite{MMV1}. In fact, these functions turn out
to be the same for the spacelike and timelike cases (this holds only
in the present large-$x$ limit, not for the full real emission contributions).
Finally the standard renormalization and mass factorization is performed to the
third order for the resulting timelike analogue of Eq.~(\ref{FSdec}), yielding
the $\DDk$ and $\delta \x1$ terms of $c_{a}^{(3)}(x)$ in Eq.~(\ref{Cqexp}). 

For the convenience of the reader, we include also the large-$x$ limits of the
well-known first- and second-order \MSb\ coefficient functions 
\cite{NLOcoeff,RvN2loop}. 
As expected from the above discussion, these and the third-order coefficient 
function share all non-$\z2$ terms with their spacelike counterparts, hence we 
will present them via the corresponding differences 
$\delta_{\,\rm TS}^{}\, c_{n}^{} = c_{\rm T}^{(n)} - c_{\rm S}^{(n)\!}$. 
The results read
\bea
\label{dTSc1}
  \delta_{\,\rm TS}^{}\, c_{1}^{}(x)  &\,=\,& 
%%START
%%L %%texdTSc1 =
       12\, \* \z2\, \* \cf\, \* \delta\x1
%%;
%%STOP
\:\: ,
\\[2mm]
\label{dTSc2}
  \delta_{\,\rm TS}^{}\, c_{2}^{}(x)  &\!=\!&
%%START
%%L %%texdTSc2 =
       48\, \* \z2\, \* \cfs \* \;\DD{1}
     - 36\, \* \z2\, \* \cfs \* \;\DD{0}
\nn\\[1mm] &&\mbox{\hspn}
     + \bigg\{ ( - 108 + 24\, \* \z2 )\, \* \cfs
       + \bigg( \, {466 \over 3} - 24\, \* \z2\ \bigg) \* \ca \* \cf
       - {76 \over 3}\: \* \cf \* \nf \bigg\}\, \* \z2\: \* \delta\x1 
%%;
%%STOP
\:\: , \quad
\\[2mm]
\label{dTSc3}
  \delta_{\,\rm TS}^{}\, c_{3}^{}(x) &\!=\!&
%%START
%%L %%texdTSc3 =
       96\, \* \z2\, \* \cft \* \;\DD{3}
     - \{ 216\, \* \cft + 88\, \* \ca \* \cfs - 16\, \* \cfs \* \nf \}\,
       \* \z2 \; \* \DD{2}
\nn\\[1.5mm] &&\mbox{\hspn}
     - \bigg\{ (324 + 96\, \* \z2 )\, \* \cft 
       - \bigg( \, {3332 \over 3} - 192\, \* \z2\ \bigg) \* \ca \* \cfs
       + {536 \over 3}\: \* \cfs \* \nf \bigg\}\, \* \z2
       \* \;\DD{1}
\nn\\[1mm] &&\mbox{\hspn}
     + \bigg\{ ( 306 + 216\, \* \z2 - 96\, \* \z3 )\, \* \cft
       - \bigg( \,{10504 \over 9} - 248\, \* \z2 - 480\, \* \z3 \bigg)
         \* \ca \* \cfs 
\nn\\[1mm] &&\mbox{}
       + \bigg( \, {1672 \over 9} - 32\, \* \z2 \bigg) \* \cfs \* \nf \bigg\}\, 
     \* \z2 \; \* \DD{0}
\nn\\[1mm] &&\mbox{\hspn}
     + \bigg\{ \bigg( \, {993 \over 2} + 180\, \* \z2 - 936\, \* \z3
         + 72\, \* \zss \bigg) \* \cft
       - \bigg(\, {13457 \over 6} + {220 \over 3}\: \* \z2 - 1616\, \* \z3
\nn\\[1mm] &&\mbox{}
       + {108 \over 5}\: \* \zss \bigg) \* \ca \* \cfs
       + \bigg(\, {74728 \over 27} - 196\, \* \z2 - 1056\,\* \z3 
         + {528 \over 5}\: \* \zss \bigg) \* \cas \* \cf
\nn\\[1mm] &&\mbox{}
       + \bigg(\, {667 \over 3} + {136 \over 3}\: \* \z2 - 80\, \* \z3
         \bigg) \* \cfs \* \nf
       - \bigg(\, {23504 \over 27} + {16 \over 3}\, \* \z2 - 96\, \* \z3
         \bigg) \* \ca \* \cf \* \nf
\nn\\[1mm] &&\mbox{}
       + \bigg(\, {1624 \over 27} + {16 \over 3}\: \* \z2 \bigg) \* \cf \* \nfs
     \bigg\}\: \* \z2\, \* \delta\x1
%%;
%%STOP
\:\: .
\eea
Here $C_A$ and $C_F$ are the standard group invariants, with $C_A = 3$ and
$C_F = 4/3$ in QCD, and $\nf$ the number of light flavours. $\zeta_k$ denotes 
Riemann's $\zeta$-function.
The third-order SIA coefficient functions can be obtained by adding the 
corresponding DIS results given in Eqs.~(4.14) -- (4.19) and Appendix B of 
Ref.~\cite{MVV6}, see also Eq.~(3.8) of Ref.~\cite{MVV10}. 
The first half of Eq.~(\ref{dTSc3}) agrees with the result of Ref.~\cite
{Blumlein:2006pj}, the $\delta\x1$ contribution in the second half has not been
presented before.
 
Below we will need the $N$-independent parts 
$ 
  \delta_{\,\rm TS}^{}\, g_{0k}^{} \: \equiv \: 
  \delta_{\,\rm TS}\, c_{k}^{}(N) \big|_{N^0}
$
of the Mellin transforms of Eq.~(\ref{dTSc1}) -- (\ref{dTSc3}) obtained via
\beq
\label{plustoN}
  a^{\,N} \; = \; \int_0^1 \! dx \, \left( x^{\,N-1} - 1 \right) a(x)_+
\eeq
together with $\delta \x1 \ra 1$. These contributions are given by
($\ec$ is the Euler-Mascheroni constant)
\bea
\label{eq:dTSg01}
  \zeta_2^{\,-1}\, \delta_{\,\rm TS}^{}\, g_{01}^{} &\!=\! & 
%%START
%%L %%texdTSg01 =
     12\, \* \cf
%%;
%%STOP
\:\: ,
\\[1mm]
\label{eq:dTSg02}
  \zeta_2^{\,-1}\, \delta_{\,\rm TS}^{}\, g_{02}^{} &\!=\!& 
%%START
%%L %%texdTSg02 =
     \ca \* \cf \* \bigg( \, {466 \over 3} - 24\, \* \z2 \bigg)
     - \cfs\, \* \left( 108 - 48\, \* \z2 - 36\, \* \ec 
     - 24\, \* \ecs\, \right)
     - {76 \over 3}\: \* \cf \* \nf
%%;
%%STOP
\:\: , \quad
\\[1mm]
\label{eq:dTSg03}
  \zeta_2^{\,-1}\, \delta_{\,\rm TS}^{}\, g_{03}^{} &\!=\!& 
%%START
%%L %%texdTSg03 =
       \cft  \*  \bigg( \, {993 \over 2} + 18\, \* \z2 - 792\, \* \z3
         + {768 \over 5}\: \* \zss - 306\, \* \ec 
         + 288\, \* \ec \* \z3 - 162\, \* \ecs
\nn\\ &&\mbox{}
         + 96\, \* \ecs \* \z2 + 72\, \* \ect + 24\, \* \ecf \bigg)
     + \ca \* \cfs  \* \bigg( - {13457 \over 6} 
         + 482\, \* \z2 + {5024 \over 3}\: \* \z3 
\nn\\ &&\mbox{}
         - {588 \over 5}\: \* \zss
         + {10504 \over 9}\: \* \ec - 160\, \* \ec \* \z2 
         - 480\, \* \ec \* \z3 + {1666 \over 3}\: \* \ecs
         - 96\, \* \ecs \* \z2
\nn\\ &&\mbox{}
         + {88 \over 3}\: \* \ect \bigg)
     + \cas \* \cf  \*  \bigg( \, {74728 \over 27}
         - 196\, \* \z2 - 1056\, \* \z3 + {528 \over 5}\: \* \zss \bigg)
\nn\\ &&\mbox{}
       + \cfs \* \nf  \*  \bigg( \, {667 \over 3} 
         - 44\, \* \z2 - {272 \over 3}\: \* \z3 - {1672 \over 9}\: \* \ec 
         + 16\, \* \ec \* \z2 - {268 \over 3}\: \* \ecs 
         - {16 \over 3}\: \* \ect \bigg)
%%STOP
\quad \nn\\ &&\mbox{}
%%START
       + \ca \* \cf \* \nf  \*  \bigg( - {23504 \over 27} 
         - {16 \over 3}\: \* \z2 + 96\, \* \z3 \bigg) 
       + \cf \* \nfs  \*  \bigg( \, {1624 \over 27} + {16 \over 3}\: \* \z2 
         \bigg)
%%;
%%STOP
\:\: .
\eea
The corresponding DIS coefficients can be found in Eqs.~(4.6) -- (4.8) of
Ref.~\cite{MVV7}.

% ----------------------------------------------------------------------------

For processes such as DIS and SIA, the dominant large-$x/\,$large-$N$ 
contributions to the \MSb\ coefficient functions $C^{\,N}$ can be resummed by a
single exponential in Mellin space \cite{SoftGlue1}
\beq
\label{eq:cNres}
  C^{\,N}(\Qs) \:\; =\:\; g_{0}^{}(\Qs) \cdot \exp\, [G^{\,N}(\Qs)] 
                     \: + \: {\cal O}(N^{-1}\ln^n N) \:\: .
\eeq
The prefactor $g_{0}^{}$ collects, order by order in the strong coupling
constant $\as$, all $N$-independent contributions. The exponent $G^{\,N}$ 
contains terms of the form $\ln^{\,k} N$ to all orders in $\as$. 
Besides the physical hard scale $\Qs$ ($\,= \mp\, \qs$ in DIS/SIA, with $q$ the
four-momentum of the exchanged gauge boson), both functions depend on the 
renormalization scale $\mu_r$ and the mass-factorization scale $\mu_{\! f}^{}$. 

The exponential in Eq.~(\ref{eq:cNres}) is build up from universal radiative 
factors for each initial- and final-state parton $p$, $\Delta_{\,\rm p}$ and 
$J_{\rm p}$, together with a process-dependent contribution $\Delta^{\rm int}$.
The resummation exponents for DIS and SIA \cite{CacCat01} take the very
similar form
\bea
\label{eq:GNdec}
  G_{\rm DIS}^{\,N} \;\; & = &
    \ln \Delta_{\,\rm q} \: + \: \ln J_{\rm q} \: + \:
    \ln \Delta^{\,\rm int}_{\,\rm DIS} \:\: , \nn \\[1mm]
  G_{\rm SIA}^{\,N} \;\; & = &
    \ln \Delta_{\,\rm q} \: + \: \ln J_{\rm q} \: + \:
    \ln \Delta^{\,\rm int}_{\,\rm SIA} 
\:\: .
\eea
The radiation factors are given by integrals over functions of the running 
coupling. Specifically, the effects of collinear soft-gluon radiation off an 
initial-state or `observed' final-state quark are collected by
\beq
\label{eq:dint}
  \ln \Delta_{\,\rm q} (\Qs,\, \mu_f^{\,2}) \:\: = \:\: \int_0^1 \! dz \,
  \frac{z^{\,N-1}-1}{1-z} \,\int_{\mu_f^{\,2}}^{(1-z)^2 \Qs}
  \frac{d\qs}{\qs}\, A(\as(\qs)) \:\: .
\eeq
Collinear emissions from an `unobserved' final-state quark lead to the 
so-called jet function,
\beq
\label{eq:Jint}
  \ln J_{\rm q} (\Qs) \:\: = \:\: \int_0^1 \! dz \,\frac{z^{\,N-1}-1}{1-z}
  \, \left[ \int_{(1-z)^2 \Qs}^{(1-z) \Qs} \frac{d\qs}{\qs}\,
  A(\as(\qs)) + B (\as([1-z] \Qs)) \right] \:\: .
\eeq
Finally the process-dependent contributions from large-angle soft gluons are 
resummed by
\beq
\label{eq:Dint}
  \ln \Delta^{\rm int} (\Qs) \:\: = \:\: \int_0^1 \! dz
  \,\frac{z^{\,N-1}-1}{1-z} \, D(\as([1-z]^2 \Qs)) \:\: .
\eeq
The functions $g_{0}^{}$ in Eq.~(\ref{eq:cNres}) and $A$, $B$ 
and $D$ in Eqs.~(\ref{eq:dint}) -- (\ref {eq:Dint}) are given by the expansions
\beq
\label{eq:aexp}
  F(\as) \:\: = \:\:
  \sum_{l=l_0}\, F_l\: \frac{\as^{\,l}}{4\pi} \:\: \equiv \:\:
  \sum_{l=l_0}\, F_l\: \ar^{\,l} \:\: ,
\eeq
where $l_0= 0$ with $g_{00}^{}= 1$ for $F= g_0^{}$, and $l_0= 1$ else.

The known expansion coefficients of the cusp anomalous dimension (the 
coefficients of $\DD{0} \equiv 1/\x1_+$ in the \MSb\ quark-quark splitting 
functions) read \cite{Kodaira:1981nh,MVV3}
\bea
\label{eq:Aqexp}
  A_{1} & \! = \! & 4\, C_F \nn \\
  A_{2} & \! = \! & 8\, C_F \bigg[ \bigg( \,\frac{67}{18}
     - \zeta_2^{} \bigg) C_A - \frac{5}{9}\,\nf \bigg] \nn \\[1mm]
  A_{3} & \! = \! &
     16\, C_F \bigg[ C_A^{\,2} \,\bigg( \,\frac{245}{24} - \frac{67}{9}
     \: \z2 + \frac{11}{6}\:\z3 + \frac{11}{5}\:\zss \bigg)
   \: + \: C_F \nf\, \bigg( -  \frac{55}{24}  + 2\:\z3
   \bigg) \nn\\ & & \mbox{} \qquad
   + \: C_A \nf\, \bigg( - \frac{209}{108}
         + \frac{10}{9}\:\z2 - \frac{7}{3}\:\z3 \bigg)
   \: + \: \nfs \bigg( - \frac{1}{27}\,\bigg) \bigg] \:\: .
\eea
The first three coefficients of the jet function (\ref{eq:Jint}) are given by
\cite{SoftGlue1,MVV2,MVV7}
\bea
\label{eq:B1}
  B_{1} & \! =\!& - 3\: \cf
\:\: ,
\\[2mm]
\label{eq:B2}
  B_{2}  &\! =\!&
  \cfs \* \bigg[ - {3 \over 2} + 12\: \* \z2 - 24\: \* \z3 \bigg]
  + \cf \* \ca \* \bigg[ - {3155 \over 54} + {44 \over 3}\: \* \z2
  + 40\: \* \z3 \bigg]
  \nn \\[1mm] & & \mbox{\hspn}
  + \cf \* \nf \* \bigg[ \, {247 \over 27} - {8 \over 3}\: \* \z2 \bigg]
\:\: ,
\\[2mm]
\label{eq:B3}
  B_{3}  &\! =\! &
       \cft  \*  \bigg[  - {29 \over 2}\: - 18\: \* \z2 - 68\: \* \z3
         - {288 \over 5}\: \* \zss + 32\: \* \z2 \* \z3
         + 240\: \* \z5 \bigg]
\nn\\[1mm]
&&\mbox{\hspn}
       + \ca \* \cfs  \*  \bigg[  - 46 + 287\: \* \z2
         - {712 \over 3}\: \* \z3 - {272 \over 5}\: \* \zss
         - 16\: \* \z2 \* \z3 - 120\: \* \z5 \bigg]
\nn\\[1.5mm]
&&\mbox{\hspn}
       + \cas \* \cf  \*  \bigg[  - {599375 \over 729}
         + {32126 \over 81}\: \* \z2 + {21032 \over 27}\: \* \z3
         - {652 \over 15}\: \* \zss - {176 \over 3}\: \* \z2 \* \z3
         - 232\: \* \z5 \bigg]
\nn\\[1.5mm]
&&\mbox{\hspn}
       + \cfs \* \nf  \*  \bigg[ \, {5501 \over 54} - 50\: \* \z2
         + {32 \over 9}\: \* \z3 \bigg]
       + \cf \* \nfs  \*  \bigg[  - {8714 \over 729}
         + {232 \over 27}\: \* \z2 - {32 \over 27}\: \* \z3 \bigg]
\nn\\[1.5mm]
&&\mbox{\hspn}
       + \ca \* \cf \* \nf  \*  \bigg[ \, {160906 \over 729}
         - {9920 \over 81}\: \* \z2 - {776 \over 9}\: \* \z3
         + {208 \over 15}\: \* \zss \bigg]
\:\: .
\eea

Together with Eqs.~(\ref{eq:dTSg01}) -- (\ref{eq:dTSg03}), all functions but
$D$ in Eqs.~(\ref{eq:dint}) -- (\ref{eq:Dint}) are known to order
$\as^{\,3}$. Consequently the first three coefficients of $D^{\:\!\rm SIA}$ can
by determined by comparing the $\as$-expansion of Eq.~(\ref{eq:cNres}) with
the fixed-order results (\ref{dTSc1}) -- (\ref{dTSc3}). This procedure yields
\beq
\label{eq:DSIA}
   D^{\:\!\rm SIA}_k \:\: = \:\: 0 \:\: 
%, \quad \mbox{ for } k = 1,\, 2,\, 3 \:\: ,
\eeq
for $k = 1,\, 2,\, 3$,
hence $\Delta_{\,\rm SIA} ^{\,\rm int} \, = \, 1$ to at least N$^3$LL accuracy.
$D_1 = 0$ was, of course, included in the NLL resummation of Ref.~\cite
{CacCat01}. However, $B_2$ was unknown at that time, and only $B_2 + D_2$ 
could be extracted from the two-loop results of Refs.~\cite{RvN2loop} alone.
 
As expected from the identity of the DIS and SIA soft-emission functions 
${\cal S}_{\:\!k}$ in Eq.~(\ref{FSdec}), there is a strong similarity between 
the respective coefficient functions also in the framework of the soft-gluon 
exponentiation -- recall that
\beq
\label{eq:DDIS}
  D^{\:\!\rm DIS}_k \:\: = \:\: 0 \:\: , \quad 
  \Delta_{\,\rm DIS} ^{\,\rm int} \, = \, 1
\eeq
was proven to all orders in $\as$ in Refs.~\cite{Forte:2002ni,Gardi:2002xm}.
We expect that such a proof can also be derived for SIA. 
For the time being assuming the all-order validity of Eq.~(\ref{eq:DSIA}), the 
difference between the SIA (timelike, T) and DIS (spacelike, S) large-$N$ 
coefficient functions exponentiates as
\beq
\label{eq:cTSNres}
  \delta_{\,\rm TS}^{}\, C^{\,N}(\Qs) \;\: =\;\: 
  \delta_{\,\rm TS}^{}\, g_{0}^{}(\Qs) \cdot \exp\, [G^{\,N}(\Qs)]
                     \: + \: {\cal O}(N^{-1}\ln^n N) 
\eeq
where, after performing the integrations in Eqs.~(\ref{eq:dint}) -- 
(\ref{eq:Dint}), the function $G^{\,N}$ takes the form
\beq
\label{eq:GNexp}
  G^{\,N}(\Qs) \:\: = \:\:
  \ln N \cdot g_1^{}(\lambda) \: + \: g_2^{}(\lambda) \: + \:
  \ar\, g_3^{}(\lambda) \: + \: \ar^{\,2}\, g_4^{}(\lambda) 
  \ldots 
\eeq
with $\lambda = \beta_0\, \ar\, \ln N$. The first three expansion coefficients
of $\delta_{\,\rm TS}^{}\,g_{0}^{}$ for $\mu_r = \mu_{\! f} = Q$ have been 
given above in Eqs.~(\ref{eq:dTSg01}) -- (\ref{eq:dTSg03}). We will address the
fourth-order coefficient below.

The functions $g_1^{}$ to $g_4^{}$ have been derived in Refs.~\cite
{SoftGlue1,av00,Catani:2003zt,MVV7}. For completeness we include these 
functions, also here restricting ourselves to choice $\mu_r =\mu_{\! f}^{} =Q$
of the scales:
\bea
  g_1^{\,\rm DIS}(\lambda) & = & 
          A_1 \*  (
            1
          - \ln(1-\lambda)
          + \lambda^{-1} \* \ln(1-\lambda) 
          )
\label{eq:g1n}
\:\: ,
\\[1mm]
  g_2^{\,\rm DIS}(\lambda) & = & 
        \bigl(
            A_1 \* \beta_1 
          - A_2
        \bigr) \* (
            \lambda
          + \ln(1-\lambda)
          )
          + {1 \over 2} \* A_1 \* \beta_1 \* \ln^2(1-\lambda)
\nn\\
&&\mbox{}
       - \bigl( A_1 \* \ec - B_1 \bigr) \* \ln(1-\lambda)
\label{eq:g2n}
\:\: ,
\\[1ex]
  g_3^{\,\rm DIS}(\lambda) & = & 
         {1 \over 2} \* (
            A_1 \* \beta_2 
          - A_1 \* \beta_1^2 
          + A_2 \* \beta_1 
          - A_3 
         ) \*  \biggl(
            1
          + \lambda
          - {1 \over 1-\lambda}
          \biggr)
\nn\\
&&\mbox{}
      +  A_1 \* \beta_1^2 \*  \biggl(
            {\ln(1-\lambda) \over 1-\lambda}
          + {1 \over 2} \* {\ln^2(1-\lambda) \over 1-\lambda}
          \biggr)
      + \biggl(
            A_1 \* \beta_2 
          - A_1 \* \beta_1^2 
         \biggr) \* \ln(1-\lambda)
\nn\\
&&\mbox{}
       +  (
            A_1 \* \beta_1 \* \ec 
          + A_2 \* \beta_1 
          - B_1 \* \beta_1 
          ) \* \biggl(
            1
          - {1 \over 1-\lambda} 
          - {\ln(1-\lambda) \over 1-\lambda}
          \biggr)
\nn\\
&&\mbox{}
       - \biggl(
            A_1 \* \beta_2 
          + {1 \over 2} \* A_1  \* (\ecs + \z2) 
          + A_2 \* \ec
          - B_1 \* \ec
          - B_2 
         \biggr) \* \biggl(
            1
          - {1 \over 1-\lambda}
          \biggr)
\label{eq:g3n}
\:\: ,
\eea
and
\bea
  g_4^{\rm DIS}(\lambda) & \! = \! & 
       - {1 \over 6} \* A_1 \* \beta_1^3 \* 
         {\ln^3(1-\lambda) \over (1-\lambda)^2}
       + {1 \over 2} \* (
         A_1 \* \beta_1^2 \* \ec
       + A_2 \* \beta_1^2 
       - B_1 \* \beta_1^2 
       ) \* {\ln^2(1-\lambda) \over (1-\lambda)^2}
       + {1 \over 2} \* (
         A_1 \* \beta_1^3  
       - A_1 \* \beta_1 \* \beta_2 
\nn\\
&&\mbox{}
       - A_1 \* \beta_1 \* (\ecs + \z2) 
       + A_2 \* \beta_1^2 
       - 2 \* A_2 \* \beta_1 \* \ec
       - A_3 \* \beta_1 
       + 2 \* B_1 \* \beta_1 \* \ec 
       + 2 \* B_2 \* \beta_1 
       ) \* {\ln(1-\lambda) \over (1-\lambda)^2}
\nn\\
&&\mbox{}
       - (
         A_1 \* \beta_1^3 
       - A_1 \* \beta_1 \* \beta_2  
       ) \* {\ln(1-\lambda) \over 1-\lambda}
       + \biggl(
         {1 \over 2} \* A_1 \* \beta_1^3 
       - A_1 \* \beta_1 \* \beta_2  
       + {1 \over 2} \* A_1 \* \beta_3
       \biggr) \* \ln(1-\lambda)
       + (
         A_1 \* \beta_1^3 
\nn\\
&&\mbox{}
       - A_1 \* \beta_1 \* \beta_2  
       - A_1 \* \beta_1^2 \* \ec
       + A_1 \* \beta_2  \* \ec
       - A_2 \* \beta_1^2 
       + A_2 \* \beta_2 
       + B_1 \* \beta_1^2 
       - B_1 \* \beta_2 
       ) \* \biggl(
            {1 \over 2} 
          - {1 \over 1-\lambda} 
\nn\\
&&\mbox{}
          + {1 \over 2} \* {1 \over (1-\lambda)^2} 
          \biggr)
       + {1 \over 2} \* \biggl(
         {1 \over 3} \* A_1 \* \beta_1^3 
       - {1 \over 6} \* A_1 \* \beta_1 \* \beta_2 
       - {1 \over 6} \* A_1 \* \beta_3 
       - {1 \over 3} \* A_1 \* (3\* \ec \* \z2 
                 + \ect + 2 \* \z3)
\nn\\
&&\mbox{}
       + A_2 \* \beta_1 \* \ec
       - A_2 \* (\ecs + \z2)
       - {5 \over 6} \* A_2 \* \beta_1^2 
       + {1 \over 3} \* A_2 \* \beta_2 
       + {5 \over 6} \* A_3 \* \beta_1 
       - A_3 \* \ec
       - {1 \over 3} \* A_4 
       - B_2 \* \beta_1 
\nn\\
&&\mbox{}
       + B_1 \* (\ecs + \z2)
       + 2 \* B_2 \* \ec 
       + B_3 
       \biggr) \* \biggl(
            1
          - {1 \over (1-\lambda)^2}
          \biggr)
       + {1 \over 3} \* \biggl(
         A_1 \* \beta_1^3  
       - 2 \* A_1 \* \beta_1 \* \beta_2  
       + A_1 \* \beta_3 
\nn\\
&&\mbox{}
       + A_2 \* \beta_2 
       - A_2 \* \beta_1^2 
       + A_3 \* \beta_1 
       - A_4 \biggr) \* \lambda
\label{eq:g4n}
\:\: .
\eea
Factors of $\beta_0 \,=\, 11/3\: C_A - 2/3\: \nf$ have been suppressed in 
Eqs.~(\ref{eq:g1n}) -- (\ref{eq:g4n}) for brevity. 
The dependence on $\beta_0$ is recovered by $A_k \ra A_k / \beta_0^{\,k}$, 
$\,B_k \ra B_k /\beta_0^{\, k}$, $\,\beta_k \ra \beta_k /\beta_0^{\,k+1}$ and 
multiplication of $g_3$ and $g_4$ by $\beta_0$ and $\beta_0^{\,2}$, 
respectively. Note that Eq.~(\ref{eq:g4n}) includes all known coefficient
of the beta function of QCD, see Ref.~\cite{beta3} and references therein.

All parameters entering Eqs.~(\ref{eq:g1n}) -- (\ref{eq:g4n}) are known except
for the four-loop cusp anomalous dimension $A_4$. The small (see below) impact 
of this quantity -- which first occurs in the $\as^5 \ln^3 N$ contribution to
 $\delta_{\,\rm TS}^{}\, C^{\,N}$ -- can
be included by a Pad\'e estimate as in Ref.~\cite{MVV7}, backed up
by a recent calculation of one Mellin moment of the fourth-order quark-quark
splitting function \cite{Baikov:2006ai}, cf.~also Ref.~\cite{av-dis07}. 
E.g., for $\nf = 5$ one may use $A_4 \approx 1550$ (recall our small expansion
parameter $\ar = \as / (4\pi)\:\!$) and assign a conservative uncertainly of
50\% to this value.

Due to the vanishing of $\delta_{\,\rm TS}^{}\, g_{00}^{}$ the two highest
logarithms, $\as^{\,l} \ln^{2l} N$ and $\as^l \ln^{2l-1} N$, are the same for 
the SIA and DIS structure functions to all orders in $\as$. The expansion of 
Eq.~(\ref{eq:cTSNres}) with Eqs.~(\ref{eq:dTSg01}) -- (\ref{eq:dTSg03}) 
provides the six highest logarithms, cf.~Ref.~\cite{MVV7}, of the coefficient-%
function difference $\delta_{\,\rm TS}^{}\, C^{\,N}$, $\as^l \ln^{2l-a} N$ with
$a = 2,\, \ldots,\, 7$, at all orders from the fourth. In particular, all 
$\ln N$ enhanced terms are thus fixed at order $\as^{\,4}$. 
After transformation to $x$-space these contributions~read
\bea
\label{dTSc4}
  \delta_{\,\rm TS}^{}\, c_{4}^{}(x) &\!=\!&
%%START
%%L %%texdTSc4 =
       96\, \* \z2\, \* \cff \* \;\DD{5}
     - \{ 360\, \* \cff + {880 \over 3}\: \* \ca \* \cft - {160 \over 3}\: 
       \* \cft \* \nf \}\, \* \z2 \; \* \DD{4}
\nn\\[1.5mm] &&\mbox{\hspn}
     - \bigg\{ (432 + 576\, \* \z2 )\, \* \cff
       - ( 3552 - 576\, \* \z2\ )\, \* \ca \* \cft
       + 576\, \* \cft \* \nf - {1936 \over 9}\: \* \cas \* \cfs
\nn\\[1mm] &&\mbox{}
       + {704 \over 9}\: \* \ca \* \cfs \* \nf 
       - {64 \over 9}\: \* \cfs \* \nfs
     \bigg\}\, \* \z2 \* \;\DD{3}
   \;\;\; + \;\;\; \bigg\{ ( 1674 + 2160\, \* \z2 + 192\, \* \z3 )\, \* \cff
\nn\\[1mm] &&\mbox{}
       - \bigg( \,{25238 \over 3} - 2800\, \* \z2 - 2880\, \* \z3 \bigg)
         \* \ca \* \cft
       + \bigg( \, {4100 \over 3} - 352\, \* \z2 \bigg) \* \cft \* \nf 
\nn\\[1mm] &&\mbox{}
       - \bigg( \, {9616 \over 3} - 528\, \* \z2 \bigg) \* \cas \* \cfs
       + \bigg( \, {3248 \over 3} - 96\, \* \z2 \bigg) \* \ca \* \cfs \* \nf
       - {256 \over 3}\: \* \cfs \* \nfs
      \bigg\}\, \* \z2 \* \;\DD{2}
\nn\\[1mm] &&\mbox{\hspn}
     + \bigg\{ \bigg( \, 1122 + 936\, \* \z2 - 4320\, \* \z3
         - {1248 \over 5}\: \* \zss \bigg) \* \cff
       - \bigg(\, {22916 \over 3} + {23120 \over 3}\: \* \z2 
\nn\\[1mm] &&\mbox{}
       - 3584\, \* \z3
       - {4368 \over 5}\: \* \zss \bigg) \* \ca \* \cft
       + \bigg(\, {488 \over 3} + {4592 \over 3}\: \* \z2 + 64\, \* \z3
         \bigg) \* \cft \* \nf
\nn\\[1mm] &&\mbox{}
       + \bigg(\, {224230 \over 9} - {17176 \over 3}\: \* \z2 - 7392\,\* \z3
         + {5184 \over 5}\: \* \zss \bigg) \* \cas \* \cfs
\nn\\[1mm] &&\mbox{}
      - \bigg(\, {69728 \over 9} - {3056 \over 3}\, \* \z2 - 576\, \* \z3
         \bigg) \* \ca \* \cfs \* \nf
       + \bigg(\, {4888 \over 9} - {64 \over 3}\: \* \z2 \bigg) \* \cfs \* \nfs
     \bigg\}\: \* \z2\, \* \;\DD{1}
\nn\\[1mm] &&\mbox{\hspn}
     - \bigg\{ \bigg( \, {3003 \over 2} + 3312\, \* \z2 - 3288\, \* \z3
         + 792\, \* \zss + 192\, \* \z2 \* \z3 - 5184\, \* \z5 \bigg) \* \cff
\nn\\[1mm] &&\mbox{}
       - \bigg(\, {24507 \over 2} + {78428 \over 9}\: \* \z2
       - 8816 \, \* \z3 - 1452\, \* \zss - 1728\, \* \z2 \* \z3
\nn\\[1mm] &&\mbox{}
       - 1440\, \* \z5 \bigg) \* \ca \* \cft 
       + \bigg(\, {6620501 \over 243} - {243752 \over 27}\: \* \z2 
         - {168560 \over 9}\: \* \z3 + {5952 \over 5}\: \* \zss 
\nn\\[1mm] &&\mbox{}
         + 1664\, \* \z2 \* \z3 
       + 2784\, \* \z5 \bigg) \* \cas \* \cfs
       + \bigg(\, {3551 \over 9} + {13568 \over 9}\: \* \z2
         + {688 \over 3}\:  \* \z3
         \bigg) \* \cft \* \nf
\nn\\[1mm] &&\mbox{}
       - \bigg(\, {1983208 \over 243} - {66392 \over 27}\, \* \z2 
         - 2336\, \* \z3 + {1152 \over 5}\: \* \zss \bigg) 
         \* \ca \* \cfs \* \nf
\nn\\[1mm] &&\mbox{}
       + \bigg(\, {135020 \over 243} - {464 \over 3}\: \* \z2 
         + {128 \over 9}\:  \* \z3\bigg) \* \cfs \* \nfs
     \bigg\}\: \* \z2\, \* \;\DD{0} 
%%;
%%STOP
     \;\;\; + \;\;\; \ldots \;\; .
\eea
The first four terms correspond to a NNLO + NLL accuracy as first obtained for 
DIS in Ref.~\cite{av99}. For the present case these terms have been presented, 
in a different notation, already in Ref.~\cite{Blumlein:2006pj}.  
The coefficients of $\DD{1}$ and $\DD{0}$ (recall the definition below 
Eq.~(\ref{Dplus})) are new results of the present study. 
The latter coefficient depends on our assumption that Eq.~(\ref{eq:DSIA})
extends to $k=4$.

% ----------------------------------------------------------------------------

The fourth-order result (\ref{dTSc4}) can be verified, and extended to the
$\delta \x1$ contribution, in the following manner. Eq.~(\ref{FSdec})
is extended to the fourth order,
\pagebreak
\bea
\label{FSdec4}
 F_{\rm S,4}^{\,\rm b}
     &\!=\!& ( 2 {\cal F}_4
           + 2 {\cal F}_1 {\cal F}_3
           + \left( {\cal F}_2 \right)^2 ) \, \delta(1-x)
           + (\, 2 {\cal F}_3 + 2 {\cal F}_1 {\cal F}_2 )\, {\cal S}_{\,1}
\nn\\[1mm]
&&\mbox{}
           + (\, 2 {\cal F}_2 + \left({\cal F}_1\right)^2 \,)\, {\cal S}_{\,2}
           + 2 {\cal F}_1 {\cal S}_{\,3} + {\cal S}_{\,4}
\:\: , 
\eea
and is subtracted from its timelike counterpart obtained as discussed above. 
Assuming that also ${\cal S}_{\,4}$ is identical in the two cases, the only
unknown in $\delta_{\,\rm TS}^{}\, F_{4}^{\,\rm b}$ to order $\ep^0$ is 
the four-loop anomalous dimension $A_4$. All other unknown quantities, such as 
the $\ep^1$ and $\ep^2$ contributions to the spacelike three-loop form factor
\cite{MVV8,MVV9,Baikov:2009bg} (also the latter new result is not needed in the
present context), drop out in this difference. Also the four-loop form factor 
is known from its exponentiation \cite{FFexp} to a sufficient accuracy in $\ep$ 
\cite{MVV8}.
The soft and virtual contributions to $\delta_{\,\rm TS}^{}\,c_{4}^{}$ are 
then extracted from the fourth-order mass factorization formula (here given in 
terms of the bare coupling)
\bea
 \label{FSTfact4}
 \delta_{\,\rm TS}^{}\, F_{4}^{\,\rm b} &\! = \! & 
     \delta_{\,\rm TS}^{}\,c_4^{} 
 \:+\: {1 \over 3}\: [ \,\b2 - P_2^{} ] \, \delta_{\,\rm TS}^{}\,a_1^{}
 \:+\: \bigg[ \,{4 \over 3}\: \b0 \b1 - {7 \over 6}\: P_1^{} \b0 
       - {2 \over 3}\: P_0^{} \b1 + {1 \over 2}\: P_0^{} P_1^{} \bigg]
         \, \delta_{\,\rm TS}^{}\,b_1^{}
\quad \nn\\[1mm]
&&\mbox{}
 \:+\: \bigg[ \,\bb03 - {11 \over 6}\: P_0^{} \bb02 + P_0^{\,2} \b0
      - {1 \over 6}\: P_0^{\,3} \bigg] \,\delta_{\,\rm TS}^{}\,d_1^{} 
 \:+\: \bigg[ \, \b1 - {1 \over 2}\: P_1^{} \bigg] 
       \,\delta_{\,\rm TS}^{}\,a_2^{}
\nn\\[1mm]
&&\mbox{}
 \:+\: \bigg[ \, 3 \bb02 - {5 \over 2}\: P_0^{} \b0 
      + {1 \over 2}\: P_0^{\,2} \bigg] \,\delta_{\,\rm TS}^{}\,b_2^{} 
 \:+\: [ \,3 \b0 - P_0 ] \,\delta_{\,\rm TS}^{}\,a_3^{}
   \; + \; \ep\mbox{-terms} \:\: .
\eea
For brevity we have suppressed the $\ep^{-3}\!$ \dots $\,\ep^{-1}$ terms which 
form a consistency check but do not provide new information. The functions
$a_n^{}$, $b_n^{}$ and $d_n^{}$ are the $\ep^1$, $\ep^2$ and $\ep^3$ 
contributions, respectively, to the $D$-dimensional coefficient functions at 
order $\as^{\,n}$, cf.~Ref.~\cite{MVV10}, and $P_n$ denotes the N$^{\:\!n}$LO
quark-quark splitting functions. In $x$-space obviously all products of these 
functions in Eq.~(\ref{FSTfact4}) have to be read as Mellin-convolutions.
 
The determination of $\delta_{\,\rm TS}^{}\,c_{4}^{}$ from Eqs.~(\ref{FSdec4})
and (\ref{FSTfact4}) reproduces the result in Eq.~(\ref{dTSc4}) 
--- hence $D^{\:\!\rm SIA }_k = D^{\:\!\rm DIS}_k \; ( = 0)$ in 
Eq.~(\ref{eq:Dint}) corresponds to $ \delta_{\,\rm TS}^{} {\cal S}_{\,k} = 0$ 
in Eqs.~(\ref{FSdec}), (\ref{FSdec4}) and their higher-order generalizations 
--- and includes the final large-$x$ coefficient, 
\bea
  \label{dTSc4del}
  \zeta_2^{-1}\,\delta_{\,\rm TS}^{}\, c_{4}^{}\Big|_{\delta \x1} &\!\!=\!\!&
%%START
%%L %%texdTSc4del =
          \bigg(
          - {7255 \over 2}\:
          - 3779\, \* \z2
          - 3816\, \* \z3
          - {13896 \over 5}\: \* \zss
          + 4080\, \* \z2 \* \z3
          + 14880\, \* \z5
\nn\\[1mm] &&\mbox{}
          + {31856 \over 105}\: \* \zts
          - 1216\, \* \zst
          \bigg) \* \cff
       +  \bigg( \,
            {191411 \over 12}\:
          + {153802 \over 9}\: \* \z2
          - 42808\, \* \z3
\nn\\[1mm] &&\mbox{}
          + {62452 \over 9}\: \* \zss
          + {8128 \over 3}\: \* \z2 \* \z3
          - {67328 \over 3}\: \* \z5
          - {102472 \over 105}\: \* \zts
          + 4064\, \* \zst
          \bigg ) \* \cft \* \ca
\nn\\[1mm] &&\mbox{\hspn}
       +  \bigg( 
          - {14817221 \over 324}\:
          - {63347 \over 3}\: \* \z2
          + {1856680 \over 27}\:\* \z3
          + {5306 \over 45}\: \* \zss
          - 2032\, \* \z2 \* \z3
\nn\\[1mm] &&\mbox{}
          + 6256\, \* \z5
          + {2584 \over 21}\: \* \zts
          - 992\, \* \zst
          \bigg) \* \cfs \* \cas
       +  \bigg( \,
            {13294462 \over 243}\:
          + {206162 \over 27}\: \* \z2
\nn\\[1mm] &&\mbox{}
          - {416032 \over 9}\: \* \z3
          - 1100\, \* \zss
          + 1936\, \* \z2 \* \z3
          + 8976\, \* \z5
          \bigg) \* \cf \* \cat
\nn\\[1mm] &&\mbox{\hspn}
       +  \bigg( \,
            {409 \over 6}\:
          - {23350 \over 9}\: \* \z2
          + 6840\, \* \z3
          - {55592 \over 45}\: \* \zss
          - {2272 \over 3}\: \* \z2 \* \z3
          + {6272 \over 3}\: \* \z5
          \bigg) \* \cft \* \nf
%
%%STOP
\quad 
%%START
\nn\\[1mm] &&\mbox{\hspn}
       +  \bigg( \,
            {706405 \over 81}\:
          + {187834 \over 27}\: \* \z2
          - {416384 \over 27}\: \* \z3
          + {6932 \over 45}\: \* \zss
          + 320\, \* \z2 \* \z3
\nn\\[1mm] &&\mbox{}
          - 1408\, \* \z5
          \bigg) \* \cfs \* \ca \* \nf
       -  \bigg(
            {2109553 \over 81}\:
          + {106168 \over 27}\: \* \z2
          - {127000 \over 9}\: \* \z3
          + 352\, \* \z2 \* \z3
\nn\\[1mm] &&\mbox{}
          - {1088 \over 5}\: \* \zss
          + 1632\, \* \z5
          \bigg) \* \cf \* \cas \* \nf
       -  \bigg(
            {3233 \over 81}\:
          + {14824 \over 27}\: \* \z2
          - {20656 \over 27}\: \* \z3
\nn\\[1mm] &&\mbox{}
          + {2464 \over 45}\: \* \zss
          \bigg) \* \cfs \* \nfs
       +  \bigg( \,
            {305917 \over 81}\:
          + {17504 \over 27}\: \* \z2
          - {8336 \over 9}\: \* \z3
          - {16 \over 5}\: \* \zss
          \bigg) \* \cf \* \ca \* \nfs
\nn\\[1mm] &&\mbox{\hspn}
       -  \bigg(
            {39352 \over 243}
          + {304 \over 9}\: \* \z2
          + {64 \over 9}\: \* \z3
          \bigg) \* \cf \* \nft
       + \bigg(
            768
          + 1920\, \* \z2
          + 896\, \* \z3
\nn\\[1mm] &&\mbox{}
          - {384 \over 5}\: \* \zss
          - 5120\, \* \z5
          \Bigg)\, \* \fl11\, \* \cf\, \* \dabcnc 
       \:+\: 3\, \* \A4
%%;
%%STOP
\:\: .
\eea
See Ref.~\cite{MVV6} for the $\fl11$ diagram class leading to the term with
%$\displaystyle\,\dabcnc \,=\, {5 \over 18}\: \nf$ in QCD.
$\dabc2/n_c \,=\, 5/18\: \nf$ in QCD. The numerical effect of this contribution
is very small and will be disregarded in the following.
 
% ----------------------------------------------------------------------------

The Mellin transform of these equations provides the
$\as^{\,4}$ prefactor $\delta_{\,\rm TS}^{}\, g_{04}^{}$ in 
Eq.~(\ref{eq:cTSNres}), and hence (up to the residual uncertainty due to $A_4$)
the seventh tower of large-$x$ logarithms from order $\as^{\,5}$ for this
difference. For $\nf=5$ quark flavours, the numerical expansion of 
$\delta_{\,\rm TS}^{}\, g_0^{}$ is given by
\beq
\label{dg0num}
  \delta_{\,\rm TS}^{}\, g_0^{}(\as) \:\: \simeq \:\: 
%  2.0944\, \as + 3.0631\, \as^{\,2} + 5.7585\, as^{\,3} 
%  + ( 13.947 + 0.1979\, A_4/1000 ) \, as^{\,4} + {\cal O}(\as^{\,5})
  2.094\, \as \left( 1 + 1.463\, \as + 2.749\, \as^{\,2} 
  + \{ 6.659 + 0.094\, A_4/1000 \} \, \as^{\,3} + \ldots\, \right)
\:\: .
\eeq
Thus the two new terms form a correction of almost 5\% at $\as = 0.12$, with
a negligible uncertainty from the missing exact value of $A_4$, and the
fourth-order contribution is less than half of the previous term for 
$\as < 0.2$. 
It is well-known that the coefficients in Eq.~(\ref{dg0num}) are due to 
$\z2$-terms (i.e., powers of $\pi^{2\,}$) from the analytic continuation of 
the form factor which are subject to a separate exponentiation (see, e.g., 
Refs.~\cite{FFexp}).
The corresponding results for the SIA and DIS cases read
\bea
\label{gTS0num}
  g_{\rm T,0}^{}(\as) &\! = \!&
  1 \:+\: 1.045\, \as \:+\: 2.266\, \as^{\,2} \:+\: 4.703\, \as^{\,3} 
    \:+\: \ldots 
\:\: , \nn \\
  g_{\rm S,0}^{}(\as) &\! = \!&
  1 \:-\: 1.050\, \as \:-\: 0.797\, \as^{\,2} \:-\: 1.056\, \as^{\,3} 
    \:+\: \ldots
\:\: .
\eea
The pattern of the corrections in Eq.~(\ref{gTS0num}) and the size of the 
$\as^{\,4}$-term in Eq.~(\ref{dg0num}) strongly suggests that the fourth-order 
contribution to $g_{\rm T,0}^{}$ amounts to less than 0.5\% for $\as = 0.12$.
 
The coefficients of the known $\ln^{\,k} N$ terms are given in Table~1 to the 
tenth order in $\as$, using the notation $c_{ka}^{}$ for the coefficient of 
$\ar^{\,k} \ln^{\,2k-a+1} N$ in $C_{\rm SIA}^{\,N}$. Hence, as in Ref.~\cite
{MVV7} for the DIS case, the coefficients of the leading (next-to-leading etc)
logarithms are denoted by $c_{k1}^{}$ ($c_{k2}^{}$ etc). The qualitative
pattern of these coefficients is similar to the DIS case (where all numbers
$c_{k,a>2}^{}$ are smaller). The higher-order coefficients rise very rapidly, 
by about an order of magnitude or more, with $a$ until $a=k- \theta_{k4}$
without showing the larger-$a$ turnover of the DIS coefficients, cf.\ Table~1
of Ref.~\cite{MVV7}. 
Indeed, the coefficient for the two cases are very similar for $a \ll k$, but 
the SIA coefficient are more than double their DIS counterparts at $a > k$ 
where the numbers are large. 

Consequently the higher-order soft plus virtual contributions are qualitatively 
similar, but larger in the timelike case. The numerical size of its resummed 
coefficient function (\ref{eq:cNres}) is illustrated in Fig.~1 for a value of 
$\as$ corresponding to LEP$\:\!$1, $s = M_Z^{\,2\,}$. 
Obviously the size of the coefficient function, as well as the relative impact 
of the new N$^2$LL and N$^3$LL corrections, increases towards lower CM energies.
Nevertheless one can conclude from Fig.~1 that the accuracy now reached for the 
dominant large-$x/\,$large-$N$ contributions should be sufficient for the 
foreseeable future.

\begin{table}[hptb]
\begin{center}
\begin{tabular}{|r||r|r|r|r|r|r|r|}\hline
   &        &        &        &        &            \\[-4mm]
$k$& $c_{k1}^{} \quad $ & $c_{k2}^{} \quad $ &
     $c_{k3}^{} \quad $ & $c_{k4}^{} \quad $ &
     $c_{k5}^{} \quad $ & $c_{k6}^{} \quad $ &
     $ c_{k7}^{}/10\; $ \\[1mm]
                                                     \hline \hline
   &         &         &         &         &         & & \\[-4mm]
 1 & 2.66667 &  7.0785 &   ---   &   ---   &   ---   &   ---  &   ---  \\[1mm]
 2 & 3.55556 & 25.6908 & 105.621 &  104.34 &   ---   &   ---  &   ---  \\[1mm]
 3 & 3.16049 & 43.3408 & 309.335 & 1016.50 &  2306.0 &   2090 &   ---  \\[1mm]
 4 & 2.10700 & 46.6020 & 514.068 & 3125.96 & 11774.1 &  23741 &   4664 \\[1mm]
 5 & 1.12373 & 36.4525 & 577.143 & 5393.82 & 32365.2 & 110255 &  29009 \\[1mm]
 6 & 0.49944 & 22.3131 & 481.110 & 6314.54 & 55037.7 & 293931 & 119399 \\[1mm]
 7 & 0.19026 & 11.1933 & 315.972 & 5515.83 & 65426.2 & 506294 & 294105 \\[1mm]
 8 & 0.06342 &  4.7503 & 170.251 & 3808.07 & 58765.0 & 618949 & 487117 \\[1mm]
 9 & 0.01879 &  1.7455 &  77.500 & 2160.26 & 41980.1 & 574684 & 589591 \\[1mm]
10 & 0.00501 &  0.5652 &  30.470 &  1035.7 & 24725.4 & 425171 & 551698 \\[1mm]
                                                     \hline
\end{tabular}
\vspace{3mm}
\caption{Numerical values of the five-flavour coefficients $c_{ka}^{}$ of the
 $\ar^{\,k} \ln^{\,2k-a+1} N$ contributions to the coefficient function 
 $C_{\rm SIA}^{\,N}$. The first six columns are exact up to the numerical 
 truncation, and the same for $F_I$, $F_T$ and $F_A$. The seventh column 
 neglects the tiny (and non-universal) $\fl11$ contributions, and uses the
 estimate $A_4 = 1550$ for the four-loop cusp anomalous dimension.}
\vspace*{-4mm}
\end{center}
\end{table}
\begin{figure}[p]
\vspace{-2mm}
\centerline{\epsfig{file=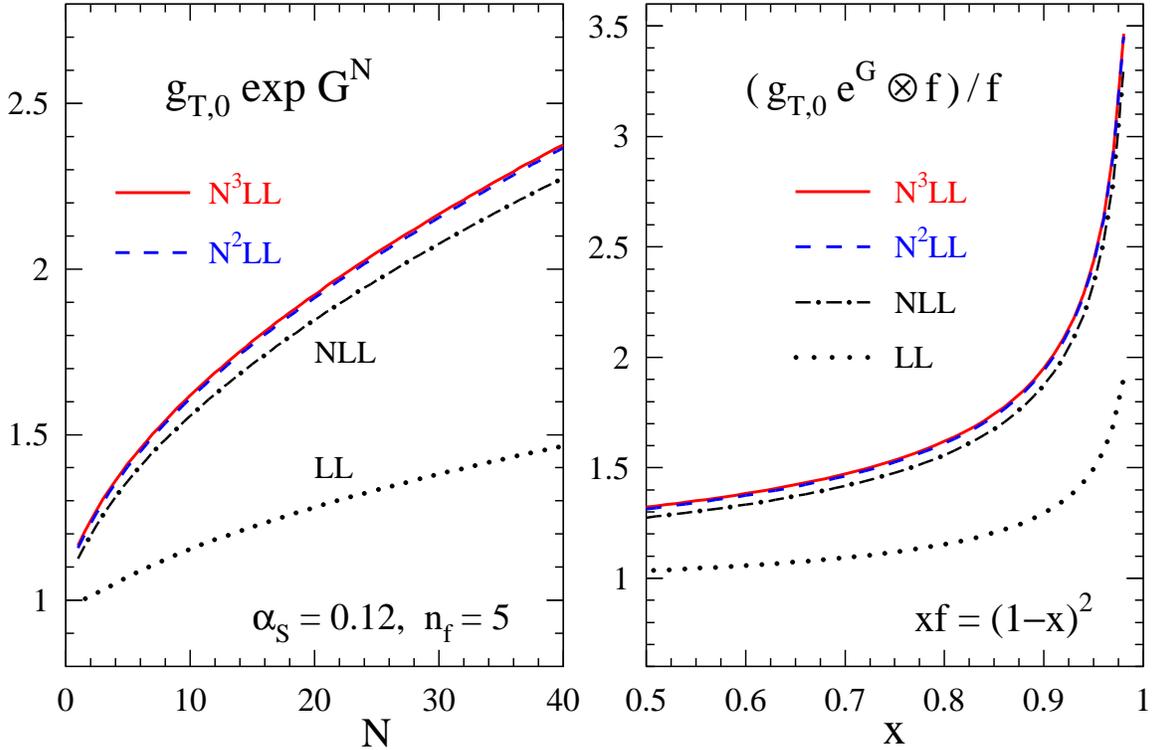,width=15.8cm,angle=0}}
\vspace{-2mm}
\caption{\label{pic:fig1}
  Left: the LL, NLL, N$^2$LL and N$^3$LL results for the threshold resummation 
  (\ref{eq:cNres}) of the SIA coefficient functions (\ref{Cqexp}) in $N$-space.
  Terms to order $\as^{\,n\,}$ are included in $g_{\rm T,0}^{}$ for the 
  N$^n$LL curves. 
  Right: the convolutions of these results with a schematic large-$x$ shape 
  for the quark fragmentation functions, using the standard `minimal 
  prescription' contour \cite{SoftGlue2} for the Mellin inversion.}
\vspace{-1mm}
\end{figure}

% -----------------------------------------------------------------------------

To summarize, we have first employed the close relation between the 
perturbative corrections to the structure functions in deep-inelastic 
scattering (DIS) and the fragmentation functions in semi-inclusive $e^+e^-$ 
annihilation (SIA), see also Refs.~\cite{STrelation}, to derive the complete 
soft and virtual corrections to the third-order quark coefficient functions 
for the latter observables.

This result then made it possible to extend the soft-gluon exponentiation in 
SIA from the next-to-leading logarithmic (NLL) contributions \cite{CacCat01} 
by two orders to N$^3$LL accuracy (we confirm the intermediate results in 
Ref.~\cite{Blumlein:2006pj}). 
It turns out that the resummation exponents are the same, presumably to all 
orders, for the DIS and SIA coefficient functions. Hence the threshold 
enhancement is structurally identical in the two cases, and the same thus holds
for the class of \mbox{large-$x$} $1/Q^{\,2}$ power corrections associated 
with the renormalon ambiguity of its perturbation series 
\cite{Gardi:2002xm,PowerCorr}.

The N$^{\,3}$LL exponentiation fixes the seven highest large-$x$ logarithms at
the fourth and all higher orders in $\as$. The especially simple connection 
between the soft and virtual contributions to the DIS and SIA coefficient 
functions also facilitates a full N$^{\,3}$LL resummation of the SIA$\,-\,$DIS 
difference, including the next-to-next-to-next-to-leading order 
$\as^{\,4}\,\delta\x1$ contribution to this difference.

Since the prefactor of the resummation exponential is larger in SIA than in 
DIS, the soft-gluon enhancement is numerically larger in the former case.
However, while the N$^{\,2}$LL contributions are still significant at LEP
energies, the N$^{\,3}$LL corrections are practically negligible, indicating
that a sufficient perturbative accuracy in the large-$x$ limit has been reached
with the present results. 

% -----------------------------------------------------------------------------
%
%\subsection*{Acknowledgments}
\vspace{5mm}
\noindent
{\bf Acknowledgments:}
Some of the symbolic manipulations for this article have been performed in 
{\sc Form} \cite{FORMref}.
S.M. acknowledges support by the Helmholtz Gemeinschaft under contract 
VH-NG-105 and in part by the Deutsche Forschungsgemeinschaft in 
Sonderforschungs\-be\-reich/Transregio~9. 
The research of A.V. has been supported by the UK Science \& Technology 
Facilities Council (STFC) under grant numbers PP/E007414/1 and ST/G00062X/1.
 
{\footnotesize
\setlength{\baselineskip}{0.5cm}

}

\end{document}